# High-*Q* silicon optomechanical microdisk resonators at gigahertz frequencies

Xiankai Sun, Xufeng Zhang, and Hong X. Tang[a)]

*Department of Electrical Engineering, Yale University, 15 Prospect St., New Haven, Connecticut 06511, USA*

**Abstract:** We report disk-shaped silicon optomechanical resonators with frequency up to 1.75 GHz in the ultrahigh frequency band. Optical transduction of the thermal motion of the disks' in-plane vibrational modes yields a displacement sensitivity of $4.1 \times 10^{-17}$ m/Hz$^{1/2}$. Due to the reduced clamping loss, these disk resonators possess high mechanical quality factors (*Q*), with the highest value of 4370 for the 1.47 GHz mode measured in ambient air. Numerical simulation on the modal frequency and mechanical *Q* for disks of varying undercut shows modal coupling and suggests a realistic pedestal size to achieve the highest possible *Q*.

[a)]Electronic mail: hong.tang@yale.edu.

Optomechanical resonators that operate at high frequencies have been a recent research focus because of their great potential both in fundamental research and in practical applications of high-speed systems for sensing and wavelength-selective signal routing.[1-4] It is well known that the frequency of a mechanical resonator can be expressed as $f = (k/m_{\text{eff}})^{1/2}/2\pi$, where $k$ denotes the modal spring constant and $m_{\text{eff}}$ the effective modal mass. Therefore the approaches to high frequencies often consist of employing high-stiffness modes, or reducing the resonator mass, or a combination of both. Achieving high mechanical quality factors ($Q$) is also very important because the transducer sensitivity and the coherence time of the mechanical vibration benefit directly from high mechanical $Q$.[5-7]

Optomechanical transduction at GHz frequencies has been demonstrated in optomechanical crystals with simultaneous bandgaps for both optical and mechanical modes. However, the construction of these structures is not trivial and requires complicated design and optimization processes.[8] On the other hand, electromechanical resonators operating in the radial-contour modes and wine-glass modes have been reported to have high mechanical $Q$ [9-11] owing to their in-plane modal vibration and immunity to air damping. Following this route, we have demonstrated optomechanical wheel resonators that operate at GHz frequencies with high mechanical $Q$ in air.[12] Those wheel structures have a large device size (radius ~30 μm) with a large resonator mass ($m_{\text{eff}}$ = 324 pg for the fundamental mode), which limits the resolution in the application of mass sensing, thus we explore microdisks with a much smaller size and greatly reduced mass to push this limit. Another advantage of smaller devices is the larger optomechanical coupling because the coupling strength, quantified as $g_{\text{om}} = \mathrm{d}\omega/\mathrm{d}R$ ($\omega$, optical resonance frequency; $R$, device radius), is expressed as $-\omega/R$ for the whispering gallery modes. Although disk-shaped optomechanical resonators were demonstrated previously, operation only in the lowest-order radial-contour mode (i.e., the radial-breathing mode) has been reported so far.[3, 13] Additionally, high mechanical $Q$ was only achieved from large disks with frequency at tens of MHz.[13] The wavelength-sized GaAs disks had low mechanical $Q$ of only around 100 due to the unoptimized undercut.[3]

In this work, by reducing the clamping loss from the disk pedestal, our silicon disk resonators vibrate at GHz frequencies with high mechanical $Q$. The disk's motion is optically transduced with help of a specially designed wrap-around coupling waveguide, which not only facilitates coupling to the disk's optical whispering gallery modes but also helps to excite and sense the disk's in-plane mechanical modes due to the symmetry.[12, 14] As a result, disk's in-plane modes at both the fundamental and higher orders with frequency up to 1.75 GHz are observed. All the modes possess high mechanical $Q$ measured in ambient air, with the highest achieved value 4370 for the 1.47 GHz mode.

Two dominant sets of in-plane mechanical modes of a disk are the radial-contour modes and the wine-glass modes. Their frequencies $f$ are determined by the following characteristic equations,[10, 11, 15, 16] under the thin-plate approximation where the disk radius $R$ is much greater than the disk thickness:

$$hRJ_0(hR)+(\sigma-1)J_1(hR)=0, \tag{1}$$

$$\left[\left.\frac{xJ_1(x)}{J_2(x)}\right|_{x=\zeta/\xi}-2-\frac{\zeta^2}{6}\right]\cdot\left[\left.\frac{xJ_1(x)}{J_2(x)}\right|_{x=\zeta}-2-\frac{\zeta^2}{6}\right]=\left(\frac{\zeta^2}{3}-2\right)^2, \tag{2}$$

where $h=2\pi f\sqrt{\rho(1-\sigma^2)/E}$, $\zeta=2\pi fR\sqrt{2\rho(1+\sigma)/E}$, and $\xi=\sqrt{2/(1-\sigma)}$. $\rho$, $\sigma$, and $E$ are the density, the Poisson ratio, and the Young's modulus of the material, respectively. $J_n$ is the Bessel function of the first kind. As seen from Eqs. (1) and (2), the frequency $f$ does not depend on the thickness of the geometry, which provides great facility in design and fabrication because devices of different resonance frequencies can easily be integrated on a chip with a single run. More importantly, the product of $f$ and $R$ is a constant for a specific mode. If the parameters $\rho = 2330$ kg/m$^3$, $\sigma = 0.27$, $E = 131$ GPa are used for single-crystal silicon, the $f{\cdot}R$ products in units of Hz·m are calculated to be 2516, 6672, 10620, … for the radial-contour modes, and 1756, 3150, 5663, … for the wine-glass modes.

Without loss of generality, we fabricated disks with a radius of 4.0 μm. The devices are fabricated from standard silicon-on-insulator substrates, with a 220-nm silicon layer on 3-μm buried oxide. The patterns are defined by electron-beam lithography and etched with chlorine-based plasma dry etching. Next, by photolithography and subsequent wet etching in a buffered oxide etchant, the oxide under the disk is mostly removed such that only a tiny pedestal remains at the center to support the disk resonator. The fabrication finishes with critical point drying. Figure 1 shows images of a device from optical microscope and scanning electron microscope (SEM). As seen from the images, a pulley-shaped wrap-around waveguide is specially designed for light to couple into and out of the disk resonator.[12, 14] Compared with the traditional point-contact coupling scheme,[17] this design not only enhances the optical coupling between the waveguide and the disk resonator, but also facilitates excitation and sensing of the in-plane mechanical modes. During the wet etching process any features in proximity of the disk resonator would also inevitably be released from the substrate, therefore a large surrounding pad with a series of triangle-shaped anchors is used to support the wrap-around waveguide to avoid falling off.

With a tunable external-cavity laser and a low-noise photodetector, the devices are prescreened by optical transmission measurement to find the coupled cavity resonances with the highest optical $Q$. The critical coupling condition is found from the devices with a gap of 80 nm between the wrap-around waveguide (width 500 nm) and the disk resonator (radius 4.0 μm). Figure 2(a) shows the transmission spectrum of such a device, displaying a good extinction ratio of more than 10 dB. Note that the background fringes result from back-reflection between two grating couplers (not shown in Fig. 1), which are used for coupling light between the on-chip waveguides and optical fibers. The measured free spectral range of the disk resonator is 22.5 nm. A narrowband transmission spectrum in the inset of Fig. 2(a) reveals a loaded optical *Q* of 70,000, corresponding to a finesse of 1020.

The mechanical modes of the disk resonator are measured by tuning the input laser wavelength to the maximum slope of an optical resonance and recording the noise spectrum of the optical transmission. The thermal motion of the disk's in-plane mechanical modes is optimally transduced because of the high optical *Q* and large optomechanical coupling.[18] The power of the laser light sent to the device is kept low such that the back-action effect is negligible and thus the measured mechanical *Q* reflects the intrinsic value. To compensate the coupling loss between the optical fibers and on-chip waveguides, the transmitted light is sent through a fiber preamplifier (Pritel FA-20) before reaching the photodetector (New Focus model 1611). The detected signal is then sent to an electrical spectrum analyzer to get the radio-frequency (RF) power spectral density.

Figure 2(b) shows the RF spectrum of the mechanical modes of the disk resonator measured in air. Since the coupling waveguide lies in the same plane and wraps around the disk resonator, only the in-plane mechanical modes can have the optomechanical coupling. The three peaks at (I) 633 MHz, (II) 1.47 GHz, and (III) 1.75 GHz are identified to be the 1st radial-contour mode, the 3rd wine-glass mode, and the 2nd radial-contour mode, respectively. Their theoretical and simulated modal frequency *f*, together with the simulated modal effective mass $m_{\text{eff}}$ and modal spring constant $k$, is summarized in Table I. While the theoretical frequencies determined from a thin-disk model are already close to the experimental values, the numerical frequencies fit even better with an underlying cone-shaped pedestal added into the simulation model. Figure 2(c)–(e) display the power spectral density zoomed in at each of the three modes, along with their modal displacement profiles from finite-element simulation. Lorentzian fit of the peaks shows that Mode I and III possess mechanical $Q$ values of around 2000, while Mode II (1.47 GHz) distinguishes itself from the others by exhibiting a significantly higher $Q$ of 4370. This remarkable difference may be attributed to their different types of motion: the wine-glass mode appears to suffer less from the clamping loss through the central pedestal than the radial-contour modes. It should be noted that the mechanical $Q$ of 4370 at the 1.47 GHz mode is about 40 times higher than that recently reported for GaAs disks at the same frequency level.[3] The $f\cdot Q$ product of $6.4 \times 10^{12}$, even though obtained in air, approaches the ultimate range of $10^{13}$, the highest yet demonstrated at room temperature in silicon mechanical resonators [6, 7] and confirmed by theoretical calculation.[19]

The power spectral density also provides a reliable way to calibrate the sensitivity of the measurement system. The spectral density of the thermomechanical displacement noise at resonance frequency is $S_x^{1/2} = \sqrt{4k_B TQ / m_{\text{eff}}(2\pi f)^3}$, where $k_B$ is the Boltzmann constant, $T$ the absolute temperature (300 K), $Q$ the mechanical quality factor, $m_{\text{eff}}$ the modal mass, and $f$ the mechanical resonance frequency.[20] By comparing the expected displacement noise with the measured spectrum, we identify the displacement sensitivity as the noise floor in the spectrum around each mode. The displacement sensitivities for Mode I and III (radial-contour modes) are similar, and at Mode III it reaches $4.1 \times 10^{-17}$ m/Hz$^{1/2}$, a value among the best that have been demonstrated in GHz optomechanical resonators and at the same order of magnitude as other sensitive nano-optomechanical systems at much lower frequencies.[3, 21] Since the sensitivity is directly related to the optical transduction efficiency, its value can vary from one mode to another because of each different optomechanical coupling strength. This may account for a lower sensitivity ($9.7 \times 10^{-17}$ m/Hz$^{1/2}$) for Mode II (wine-glass mode), whose spectrum is taken at the same optical power and detuning conditions as Mode I and III.

The RF power spectra have shown that different types of modes have quite different mechanical *Q*, indicating the clamping loss through the pedestal support as the dominant loss channel. To confirm this conjecture and further understand the influence of the pedestal, we simulated in HiQLab[22] 4.0-μm-radius disks with different underlying pedestal sizes, which correspond to devices with different undercut levels in fabrication. As illustrated in the inset of Fig. 3, the layer structure follows the wafer used for device fabrication. To set the boundary condition, a 1-μm-thick perfectly matched layer (PML) is applied to the bottom and side of the substrate, while all the other surfaces are free to move. Figure 3 plots the modal frequency and *Q* value as a function of the relative undercut $u/R$. As shown in the frequency curves, the presence of the pedestal breaks the vertical structural symmetry and leads to the coupling between the 1st radial-contour mode and the high-order radial-flapping modes. It is this coupling between the in-plane (radial-contour) modes and out-of-plane (radial-flapping) modes that results in a low clamping-loss-limited *Q* value, even at relatively high undercut levels. Checked with SEM, the disk resonator used in the measurement is found to have a relative undercut of around 0.860. A disk with this relative undercut is predicted by the simulation to have a mechanical *Q* of 1520 for the 1st radial-contour mode, which is in very good agreement with the measured value of 1600. Furthermore, the simulated mechanical *Q* curve also provides guidance for the targeted undercut level to achieve the highest possible *Q*. It is understood that in bulk-mode resonators, the silicon material-loss-limited $Q$ is estimated to be 70,000 at the frequency of 700 MHz.[19] Applying this criterion to the $Q$ curves, only by making $u/R$ either between 0.822 and 0.835 or larger than 0.945 (grayed regions in Fig. 3) can we enter the material-loss-limited regime and achieve the highest possible $Q$. For 4.0-μm-radius disks, having a relative undercut larger than 0.945 is challenging in fabrication but controlling it between 0.822 and 0.835 is feasible and realistic.

In conclusion, we have demonstrated silicon optomechanical disk resonators, vibrating in the radial-contour modes and wine-glass modes in the ultrahigh-frequency band (633 MHz, 1.47 GHz, and 1.75 GHz). The mechanical $Q$ values are around 2000 for the radial-contour modes and 4370 for the wine-glass mode, measured in ambient conditions. These modes are optically resolved by high-finesse ($F$ = 1020) cavity resonances, with the best achieved displacement sensitivity of $4.1 \times 10^{-17}$ m/Hz$^{1/2}$. Numerical simulation of disks with varying undercut reveals the coupling between the 1st radial-contour mode and the radial-flapping modes. As a result, a relative disk undercut between 0.822 and 0.835 is suggested in fabrication to achieve a material-loss-limited $Q$.

We acknowledge funding from DARPA/MTO ORCHID program through a grant from the Air Force Office of Scientific Research (AFOSR). H.X.T. acknowledges support from the National Science Foundation CAREER award and a Packard Fellowship in Science and Engineering. The authors thank Michael Power and Dr. Michael Rooks for assistance in device fabrication.

Figure captions:

FIG. 1. (Color online) (a) Dark-field optical microscope image of a device. (b) Zoomed-in image showing the disk resonator, the wrap-around coupling waveguide, and the pad used to support the suspended waveguide. (c) Tilted-view scanning electron microscope (SEM) image of the disk resonator.

FIG. 2. (Color online) (a) Optical transmission spectrum and (b) RF power spectral density (PSD) of the optical transmission of the disk resonator. (c)–(e) Zoomed-in displacement noise power spectral density of each mode showing the mechanical $Q$. Inset: normalized disk modal displacement profile.

FIG. 3. (Color online) Simulated modal frequency (lower panel) and clamping-loss-limited mechanical $Q$ (upper panel) of silicon disk with 4.0 μm radius and varying undercut. The grayed areas denote the material-loss-limited regime where the highest possible mechanical $Q$ can be achieved. Inset: cross-sectional view of the model used for simulation. PML: perfectly matched layer.

TABLE I. Properties of the observed modes, theoretical (superscript *t*), simulated (superscript *s*), and experimental (superscript *e*), of the disk resonator with radius $R$ = 4.0 μm.

| Mode No. | Mode type | $f^{(t)}$(MHz) | $f^{(s)}$ (MHz) | $f^{(e)}$ (MHz) | $m_{\mathrm{eff}}^{(s)}$ (pg) | $k^{(s)}$ ($10^6$ N/m) | $Q_{\mathrm{m}}^{(e)}$ (air) |
|---|---|---|---|---|---|---|---|
| I | radial-contour, 1st | 629.0 | 636.5 | 632.9 | 17.2 | 0.275 | 1600 |
| II | wine-glass, 3rd | 1416 | 1461 | 1470 | 5.26 | 0.443 | 4370 |
| III | radial-contour, 2nd | 1668 | 1753 | 1751 | 7.30 | 0.886 | 2040 |

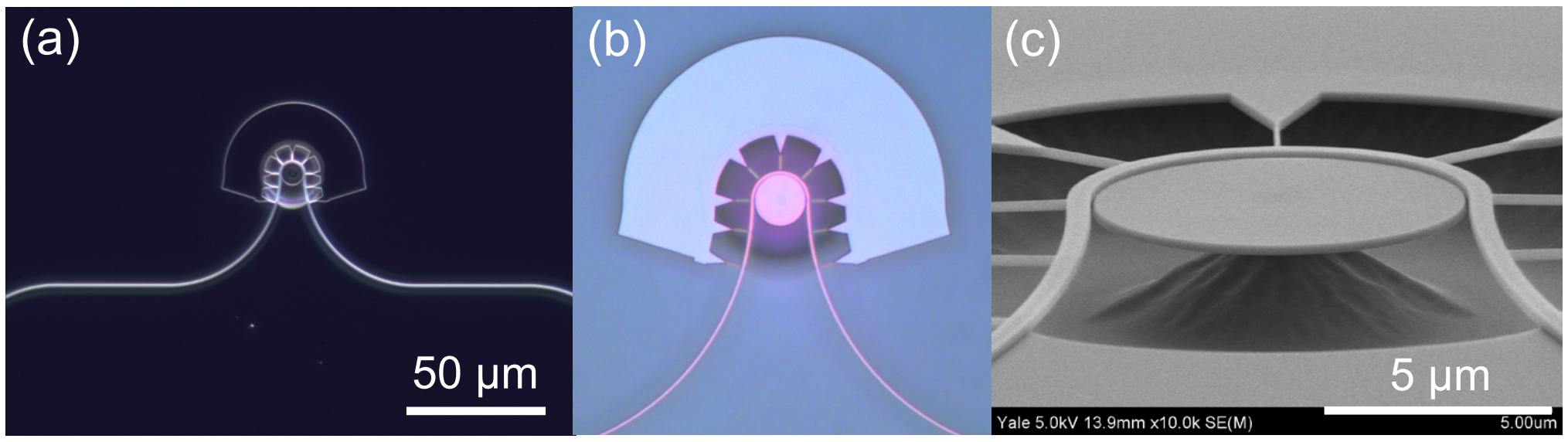

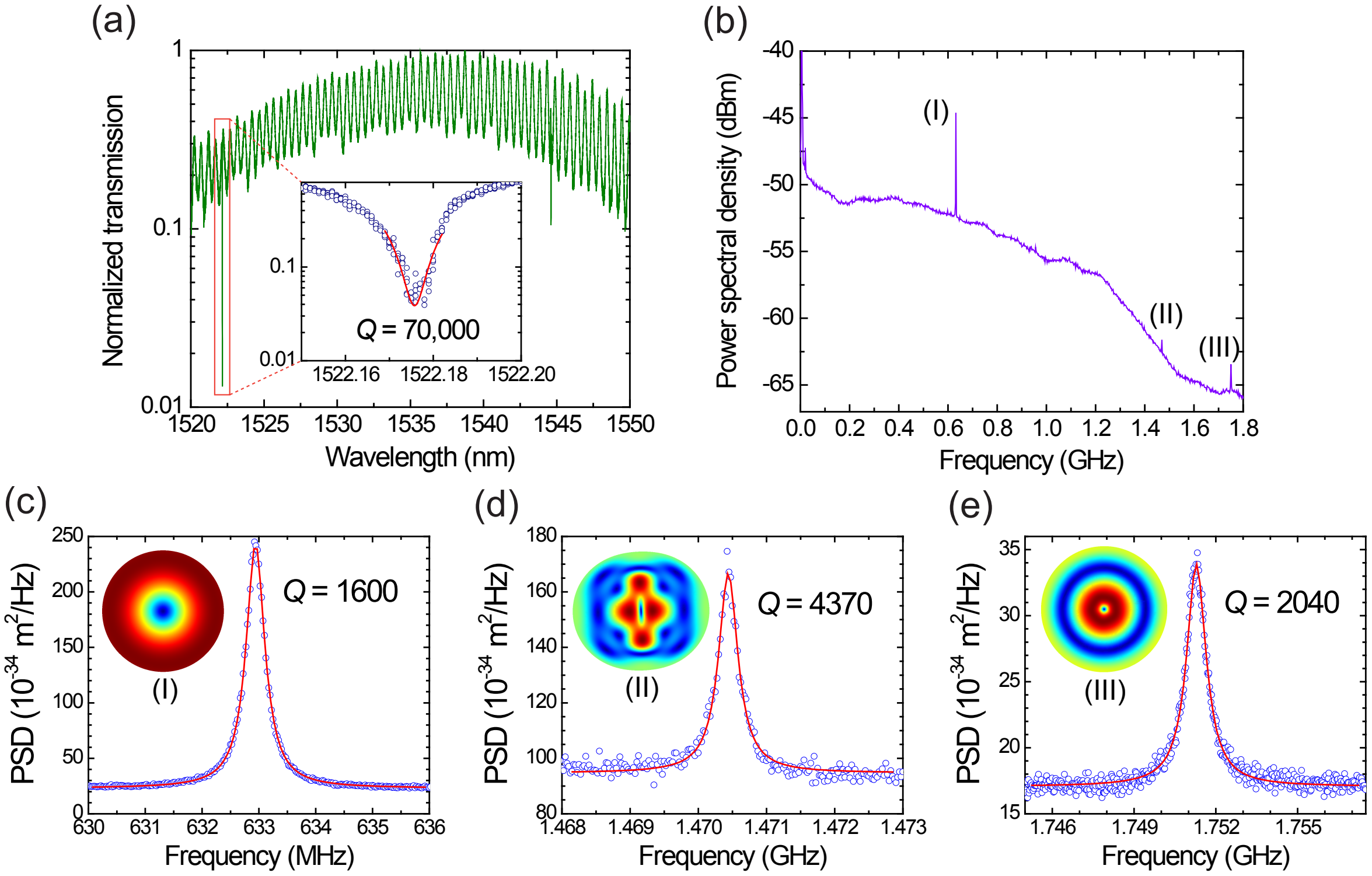

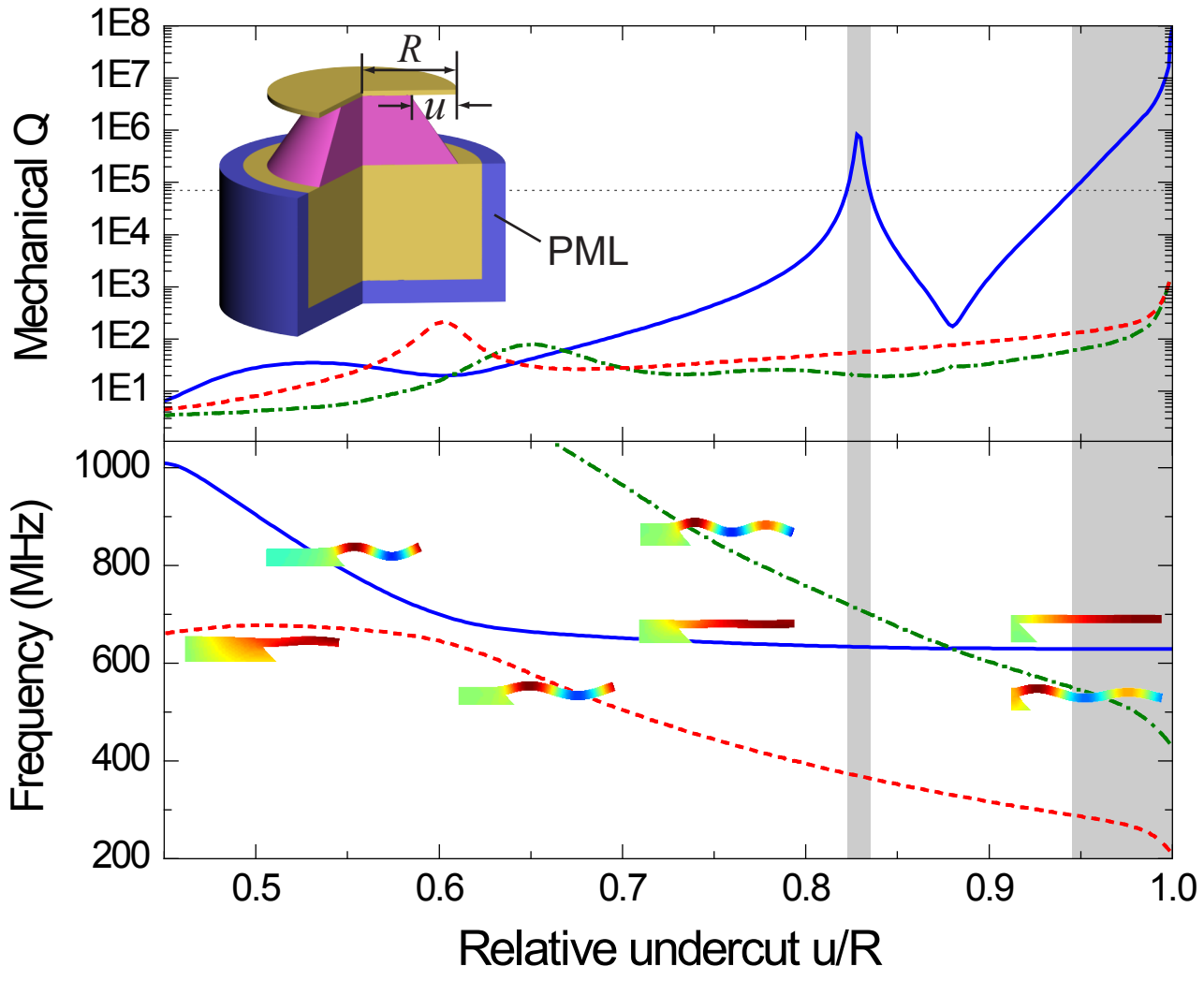